\documentclass[11pt]{article}
\usepackage{moriond,epsfig}

\def\Journal#1#2#3#4{{#1} {\bf #2}, #3 (#4)}

\def\PLB{{\em Phys. Lett.}  B}
\def\PRL{\em Phys. Rev. Lett.}
\def\PRD{{\em Phys. Rev.} D}

\def\be{\begin{equation}}
\def\ee{\end{equation}}
\def\bea{\begin{eqnarray}}
\def\eea{\end{eqnarray}}
\def\nuebar{\bar{\nu_e}}

\begin{document}
\title{HIGH SENSITIVITY ANTI-NEUTRINO DETECTION BY KAMLAND}

\author{S.\,HATAKEYAMA
\\ on behalf of the KamLAND Collaboration}

\address{Department of Physics and Astronomy, Louisiana State University,\\
Baton Rouge, Louisiana 70803, USA}

\maketitle

\abstracts{
Electron anti-neutrinos ($\nuebar$) from nuclear power reactors have been
observed by KamLAND. Data from 0.16 kton$\cdot$year exposure (145.1 live days)
indicates disappearance of $\nuebar$ at 99.95\% C.L.
in the energy range 2.6 MeV $< E_{\nuebar} <$ 8.0 MeV.
Considering two-flavor neutrino oscillation with CPT invariance, the only
remaining solution to the solar neutrino problem is the
Large Mixing Angle (LMA) solution.
In addition a 0.28kton$\cdot$year exposure (185.5 live days) was searched
for $\nuebar$ in the energy range 8.3 MeV $< E_{\nuebar} <$ 14.8 MeV.
No candidate events were found
with expected background of 1.1$\pm$0.4 events. Assuming that the origin of 
$\nuebar$ in this energy region comes from $^8$B solar $\nu_e$,
we find an upper limit of $\nu_e$ to $\nuebar$ conversion
probability of 2.8 $\times 10^{-4}$.
}

\section{KamLAND Experiment}

\subsection{Motivation of the Experiment}

Kamioka Liquid scintillator Anti-Neutrino Detector (KamLAND) started
operation on January 21st 2002.
The observation of $\nuebar$'s from nuclear power reactors was 
the primary object of the KamLAND experiment because it was expected to solve
the solar neutrino problem
and set a strong constraint on the neutrino oscillation parameters. 
The search for solar anti-neutrinos was undertaken to examine
two types of theoretical framework,
spin-flavor precession and neutrino decay.
This letter describes the reactor $\nuebar$ observation \cite{kl-1st} and
the solar $\nuebar$ search.\cite{kl-2nd}
KamLAND may also observe other neutrino sources.
The observation of geo-neutrinos in high statistics will reveal
quantitative information of U and Th in the earth.
Observation of $^7$Be solar neutrinos will be the next goal of KamLAND,
however we still need to purify the liquid scintilator to achieve a
detection of the $^7$Be neutrino signal.
Other possible neutrino sources, such as supernova neutrinos and
relic neutrinos are also an exciting prospect.
KamLAND also has the ability to detect nucleon decay.

\subsection{Detector Overview}

KamLAND is located  1000 m under ground in the Kamioka mine in Japan.
The detector is composed of a sphere shaped inner detector (ID) and a
cylindrical shaped outer detector (OD).
The inner detector which consists of 1200 m$^3$ of liquid scintillator (LS)
(90\% dodecane, 20\% pseudocumene and 1.52 g/l of PPO)
is contained within a 13 m diameter balloon
(134 $\mu$m thick transparent nylon/EVOH composite film)
which sits inside 1800 m$^3$ of buffer oil
(50\% dodecane, 50\% isoparaffin).
On the inside edge of ID sphere 1325 fast timing 17 inch diameter PMTs
and 554 20 inch diameter PMTs collect photons from the LS.
Only the 17 inch PMTs were used for this analysis.
The photo coverage was about 22\%.
The OD is filled with purified water and is instrumented with
225 20 inch diameter PMTs.
The OD is a water cherenkov detector used to eliminate cosmic ray muons.
It also acts as a shield against 
radioactivity from the surrounding rock.
KamLAND is now stable and taking data 24 hours/days except for
calibration runs or unexpected shutdowns.
The average trigger rate is $\sim$ 30Hz with the primary trigger
threshold (200 PMT hits, $\sim$0.7MeV).

\subsection{Energy and Vertex Calibration}
To calibrate the energy scale from 1MeV to several MeV, 
various radioactive sources are deployed along vertical-axis of detector.
The sources are $^{68}$Ge (1.012MeV $\gamma+\gamma$),
 $^{60}$Co (2.506MeV $\gamma+\gamma$),
 $^{65}$Zn (1.116MeV $\gamma$)
 and $^{241}$Am/Be (2.20, 4.40, 7.6MeV $\gamma$).
We also use $\gamma$-rays from neutron capture on proton (2.2MeV)
and neutron capture on $^{12}$C (4.95MeV).
These neutrons are generated by spallation from the passage of
cosmic-ray muons in the LS.
The estimated energy resolution is about 7.5\%$/\sqrt{E(\mbox{MeV})}$.
Light yield is 300 p.e./MeV.
Radioactive sources are also used to obtain the positioning bias of
the vertex reconstruction.
The bias of the reconstructed vertex position was less than
$\pm$5cm along z-axis within the fiducial volume.

\subsection{Detection of $\nuebar$ by Delayed Coincidence}

Electron anti-neutrino is detected using the inverse $\beta$ decay
reaction in the LS, 
\be
\nuebar + p \rightarrow e^+ + n.
\label{eq-inverse-beta-decay}
\ee
The generated positron is immediately annihilated to 2 $\gamma$ generating
a prompt signal. On the other hand the neutron is thermalized in
the LS and captured on a proton after $\sim$200 $\mu$s,
generating a 2.2 MeV $\gamma$ as a delayed signal.
By requiring the coincidence of prompt and delayed signal,
background can be reduced dramatically.
The visible energy of prompt signal is 
related to the $\nuebar$ energy by:
\bea
E_{\mbox{vis}}  & = & E_{\nuebar} - (\Delta m_{np} + m_e)
 - T_n(\theta) + 2m_e\\
 & = & E_{\nuebar} - 0.782\mbox{MeV} - T_n(\theta)
\label{eq-nuebar-energy}
\eea
where $ E_{\nuebar}$ is energy of the $\nuebar$,
$\Delta m_{np}$ is the mass difference between the neutron and proton,
$m_e$ is the electron mass and $T_n(\theta)$ is the kinetic energy of
the neutron scattered by an angle $\theta$.
The detection efficiency for inverse $\beta$ decay events is estimated from 
Monte Carlo simulation and calibration data to be 84.2 $\pm$ 1.5\%.
The main contributions to the detection inefficiency are the cuts on
the distance between the prompt and delayed vertices (89.8 $\pm$ 1.6\%),
the time between the prompt and delayed vertices (95.3 $\pm$ 0.3\%),
neutron capture on protons (99.5\%), and the energy of the delayed event
(98.9 $\pm$ 0.1\%).

\subsection{Spallation Events after Muon}
Although KamLAND is located deep underground (2700 m water equivalent), muons 
originating from cosmic-rays pass through the detector at a frequency
of 0.34Hz. An energetic muon can destroy a carbon nucleus in
LS by spallation. 
Various radioactive isotopes and neutrons are generated by muon spallation.
Table \ref{table-spallation-isotopes} shows a list of generated isotopes.
\begin{table}[h]
\begin{center}
\caption{Radioactive isotopes generated by muon spallation in the liquid
scintillator.}
\vspace{2mm}
\begin{tabular}{|c|c|c|} \hline
Isotope & $T_{1/2}$ & $E_{\mbox{max}}$ (keV) \\ \hline
$^{12}$B & 20.2 ms & 13369 ($\beta^-$) \\ \hline
$^{12}$N & 11.0 ms & 17338 ($\beta^+$) \\ \hline
$^{11}$Li & 8.5 ms & 20610 ($\beta^-$) \\ \hline
$^{9}$Li & 173.8 ms & 13606 ($\beta^-$, n) \\ \hline
$^{8}$He & 119.0 ms & 10653 ($\beta^-$, n) \\ \hline
$^{9}$C & 126.6 ms & 16498 ($\beta^+$) \\ \hline
$^{8}$Li & 838.0 ms & 16006 ($\beta^-$) \\ \hline
$^{6}$He & 906.7 ms & 3508 ($\beta^-$) \\ \hline
$^{8}$B & 770.0 ms & 17979 ($\beta^-$) \\ \hline
\end{tabular}
\label{table-spallation-isotopes}
\end{center}
\end{table}

Spallation events due to muons are an important source of background.
Most of the long life spallation products emit a single $\beta$-ray
so they are eliminated by the delayed coincidence requirement.
However $^9$Li and $^8$He produce both a $\beta^-$ and neutron.
For these we applied the following criteria to cut the correlated spallation
events.
(1) 2 ms veto is applied after any muon events.
(2) Additional 2 s veto is applied for energetic muon events when
the ionization energy deposit is larger than 10$^6$ p.e. ($\sim$3 GeV).
(3) For smaller energy deposits (less than 10$^6$ p.e.)
a 2 s veto is applied to events with vertices in a 3 m cylinder
around the muon track.

Muon spallation in the rock surrounding the KamLAND detector can generate
fast neutrons that penetrate through the water of outer detector.
These fast neutrons can cause a correlated background since they 
can generate both prompt and delayed signal in the LS via a
recoil proton and neutron capture $\gamma$-ray.
Adopting a 5m radius fiducial volume eliminates most of this background.

Muon spallation provides not only background but also important
calibration sources. Neutron captured on proton (2.2MeV $\gamma$),
neutron captured on $^{12}$C (4.9MeV $\gamma$) 
and $^{12}$B decayed $\beta$ are used for energy and vertex calibrations.

\vspace{4mm}

\section{Result of Reactor $\nuebar$ Observation}

KamLAND is the first experiment to observe $\nuebar$ disappearance
from a reactor source. \cite{kl-1st}
In this section the reactor experiment analysis is described.
The dominant background of reactor $\nuebar$'s is
geo-$\nuebar$'s from $\beta$ decay of U and Th
in the earth. 
The geo-$\nuebar$ flux has a expected energy distribution
from 1 MeV to 2.4 MeV. \cite{geonu}
Since the total flux of geo-$\nuebar$'s has large uncertainty,
we applied an analysis energy threshold at 2.6 MeV for the reactor analysis.

\subsection{$\nuebar$ Flux from Reactors}

Nuclear reactor power plants are strong electron anti-neutrino source.
$\nuebar$'s are generated by $\beta$ decay
of daughter nuclei from the fission of various fuel components,
such as $^{235}$U, $^{238}$U, $^{239}$Pu and $^{241}$Pu.
The flux of $\nuebar$ from a reactor is calculated from the thermal power
generation and the distance between the detector and reactor.
The thermal power information is provided by the power companies
with better than 2\% accuracy.
About 70GW (7\% of the world total) of reactor power is generated 
at a distance of 175$\pm$35km from Kamioka.
This corresponds to 80\% of $\nuebar$ flux at KamLAND.
Fortunately this distance is suitable to observe LMA neutrino oscillation.  
In our data set of 145.1 live days,
the expected number of reactor $\nuebar$ events in the fiducial volume
was 86 $\pm$ 5.6 events.
The systematic uncertainties for reactor $\nuebar$ detection are
listed in the table \ref{table-reactor-systematics}.

\begin{table}[h]
\begin{center}
\caption{Systematic uncertainties of reactor $\nuebar$ (\%).}
\vspace{2mm}
{\small
\begin{tabular}{|lc|lc|} \hline
Total LS mass & 2.1 & Reactor power & 2.0\\
Fiducial mass ratio & 4.1 & Fuel composition & 1.0\\
Energy threshold & 2.1 & Time lag of $\beta$ decay & 0.28\\
Efficiency of cuts & 2.1 & $\nu$ spectra & 2.5\\
Live time & 0.07 & Cross section & 0.2\\
\hline
Total systematic error & & & 6.4\\
\hline
\end{tabular}
}
\label{table-reactor-systematics}
\end{center}
\end{table}

\subsection{Event Selection for Reactor $\nuebar$}

The selection criteria for $\nuebar$ events is
(1) total charge cut, less than 10,000 p.e.
 ($\sim$30MeV) with no OD veto signal
and muon spallation cut.
(2) fiducial volume cut ($R < 5$m),
(3) time correlation cut (0.5$\mu$s $< \Delta T <$ 660$\mu$s),
(4) vertex correlation cut ($\Delta R < $1.6m),
(5) delayed energy window cut (1.8MeV $< E_{delay} <$ 2.6MeV), and
(6) cut on the delayed vertex position more than 1.2m from central vertical
axis to eliminate background from thermometers of LS.

After applying our cuts and a 2.6MeV analysis threshold,
54 events remained. The total background was estimated to be 0.95$\pm$0.99
events, where accidental background is $0.0086\pm0.0005$, background from
$^9$Li/$^8$He is $0.94\pm0.85$ and fast neutron is less than 0.5 events.
The ratio of the number of observed reactor $\nuebar$  events
to expected events without oscillation is
\be
\frac{N_{obs} - N_{BG}}{N_{expected}} =
 0.611 \pm 0.085(\mbox{stat}) \pm 0.041(\mbox{syst}).
\label{eq-ratio}
\ee
This result indicates $\nuebar$ disappearance with 99.95\% C.L.

\subsection{Interpretation with Neutrino Oscillation}

\begin{figure}[h]
\begin{center}
\epsfig{figure=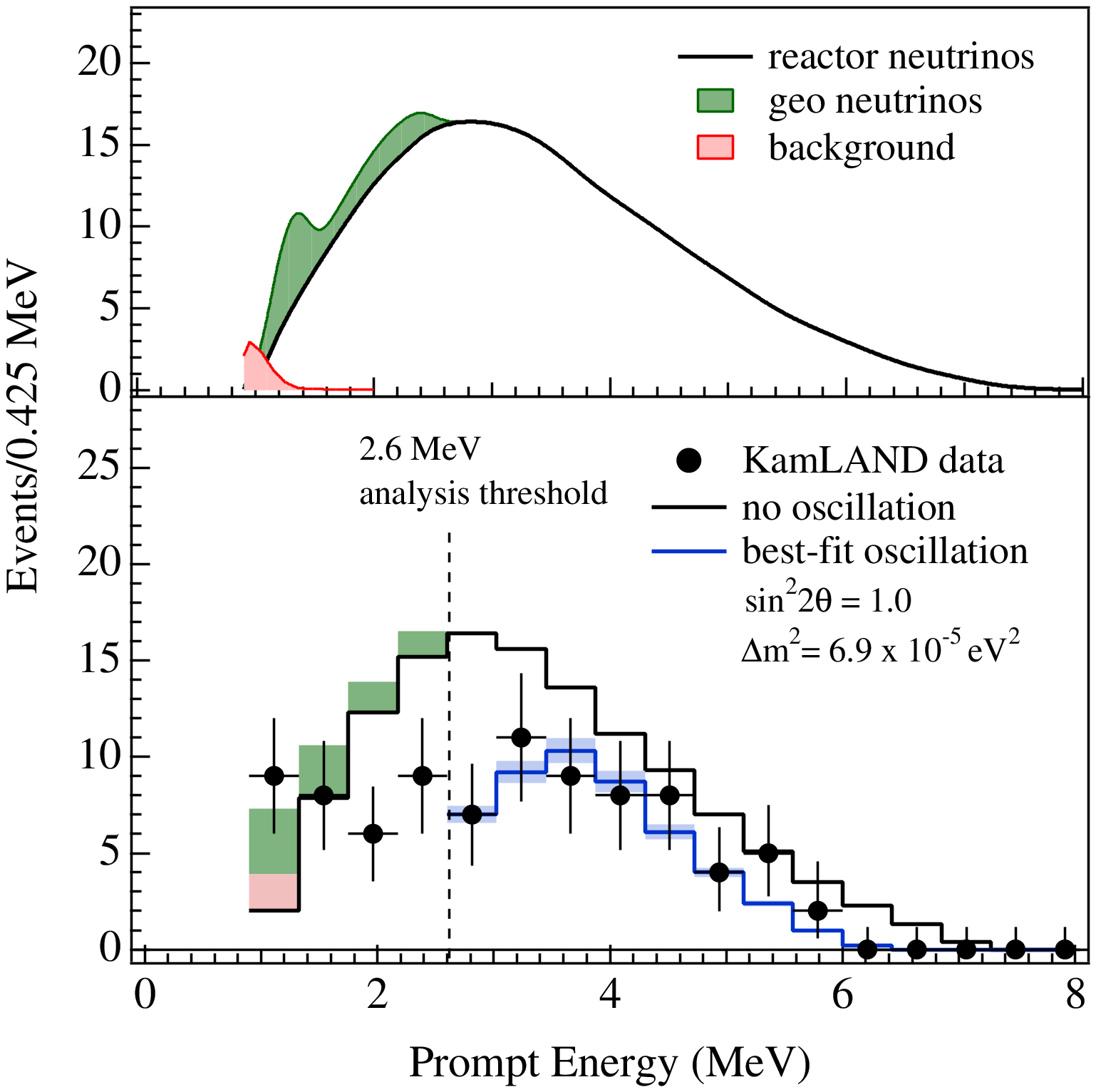,height=6.8cm,width=7.8cm}
\vspace{-2mm}
\caption{Expected $\nuebar$ spectrum from the reactor\,(upper) and
visible energy spectrum of the prompt signal\,(lower).}
\label{fig-spectrum}
\vspace{4mm}
\epsfig{figure=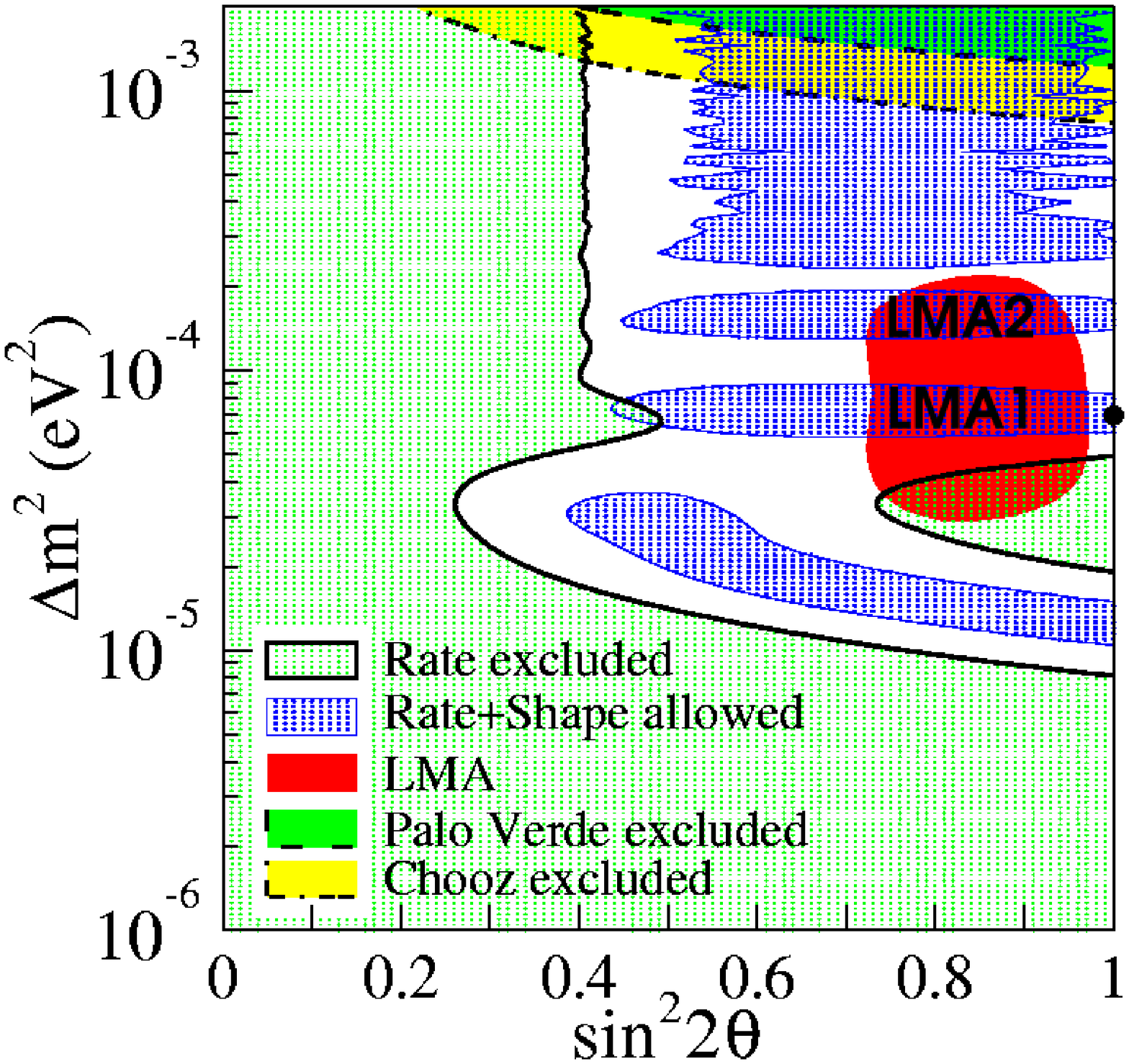,height=8.5cm,width=8.5cm}
\vspace{-2mm}
\caption{Neutrino oscillation parameters excluded by Chooz,
Palo Verde, KamLAND (Rate) and allowed by solar-$\nu$ combined (LMA),
KamLAND (Rate+Shape). Two regions (LMA-1, LMA-2) remain at 95\%.C.L.}
\label{fig-oscillation-contour}
\end{center}
\end{figure}

Neutrino oscillation is one of the most probable explanations
to understand the observed deficit of $\nuebar$'s.
A ``Rate analysis'' was performed where
oscillation parameters are examined by defining a $\chi^2$,
\be
\chi^2_{rate} = \frac{(0.611 - R(\sin^2 2\theta,\Delta m^2))^2}
{0.085^2 + 0.041^2}
\ee
where, $R(\sin^2 2\theta,\Delta m^2)$ is the ratio of expected number of
events with oscillation to expected number of events without oscillation.
Also a ``Shape analysis'', comparing the normalized energy spectrum of
$\nuebar$ was performed.
Figure \ref{fig-spectrum} shows the visible energy spectrum of the 
prompt signal that corresponds to the $\nuebar$ spectrum.
The upper figure is the expected reactor $\nuebar$ energy spectrum with
contributions from geo-$\nuebar$ (model Ia) \cite{geonu}
and accidental background. The lower figure shows the energy spectrum
of the observed
prompt events (solid circles with error bars), along with the expected no 
oscillation spectrum (upper histogram, with geo-$\nuebar$ and accidentals
shown) and best fit (lower histogram) including neutrino oscillations.
The shaded band indicates the systematic error in the best-fit spectrum.
The vertical dashed line corresponds to the analysis threshold at 2.6MeV.

A maximum likelihood function was used to define the combined $\chi^2$ of 
the rate and shape analysis.
\be
\chi^2_{rate+shape} = \chi^2_{rate}
+ \chi^2_{BG}(N_{BG1\sim2}) + \chi^2_{dist}(\alpha_{1\sim4})\\
 - 2\log L_{shape}(\sin^2 \theta, \Delta m^2, N_{BG1\sim2},\alpha_{1\sim4})
\ee
where $L_{shape}$ is the likelihood function of the spectrum including
deformations from various parameters. $N_{BG1\sim2}$ are the estimated number
of $^9$Li and $^8$He backgrounds and $\alpha_{1\sim4}$ are the parameters for
the shape deformation coming from the energy scale, resolution, $\nuebar$
spectrum and fiducial volume.

Figure \ref{fig-oscillation-contour} shows the neutrino oscillation parameter
region for two neutrino mixing at the 95\% C.L.
The excluded region from Chooz \cite{chooz} and Palo Verde \cite{palo}
are shown at the top region.
The allowed region of the LMA solution from
solar neutrino experiments \cite{solar-lma} is shown in the middle region.
The KamLAND ``Rate'' analysis  excluded most of the region except LMA.
The allowed region of LMA is split into two regions
by the KamLAND ``Rate+Shape'' analysis.
The remaining lower region is called LMA-1 and upper one is called LMA-2.
\vspace{-2mm}
\vspace{-2mm}
\subsection{Future Prospects of the KamLAND Reactor $\nuebar$ Observation}
\begin{figure}[h]
\begin{center}
\epsfig{figure=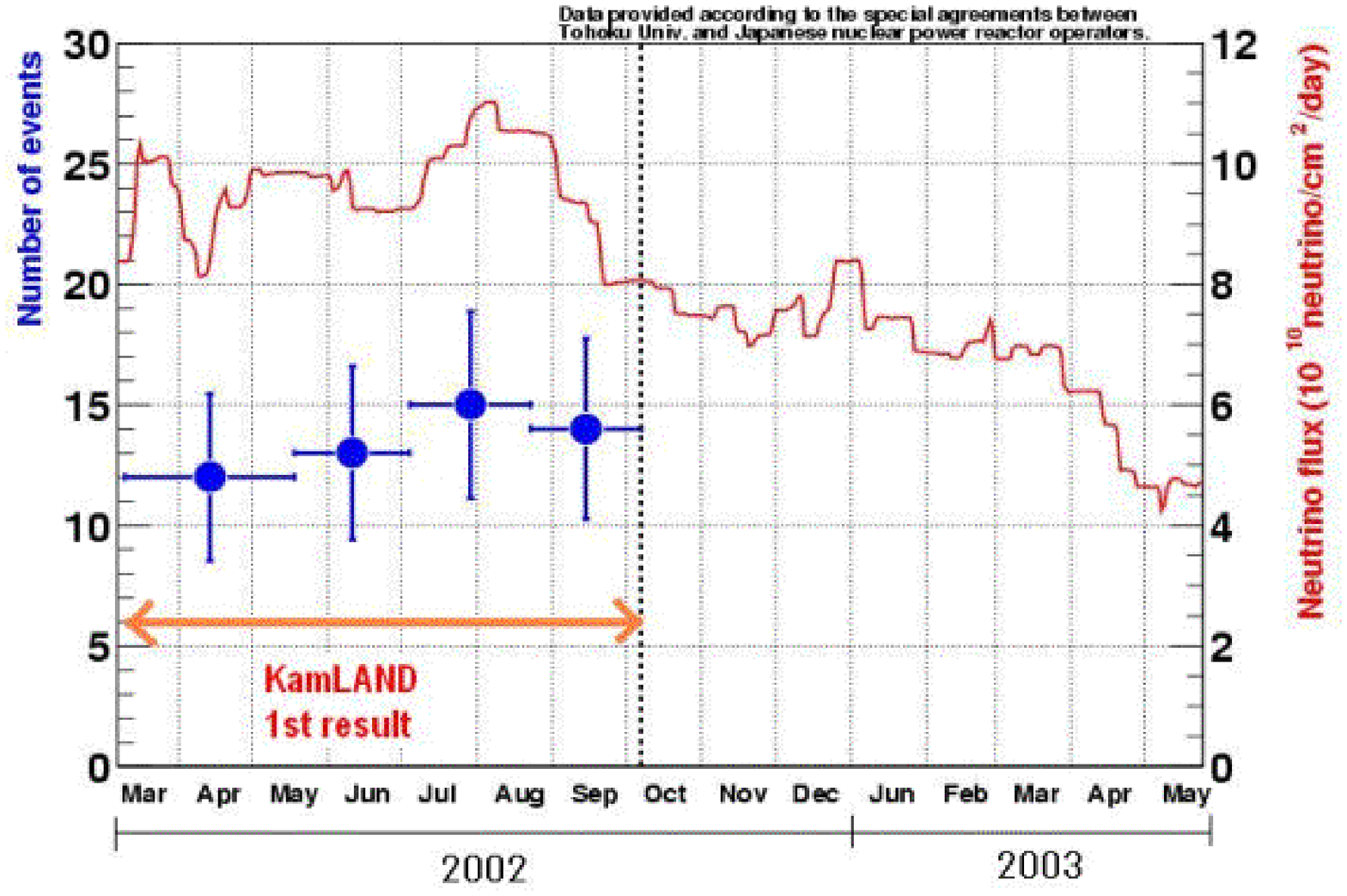,height=6.4cm}
\vspace{-2mm}
\caption{Time variation of the expected reactor $\nuebar$ flux
at KamLAND is shown (solid line). The observed number of reactor
$\nuebar$ events is also shown (solid circles with error bars).}
\label{fig-modulation}
\vspace{2mm}
\epsfig{figure=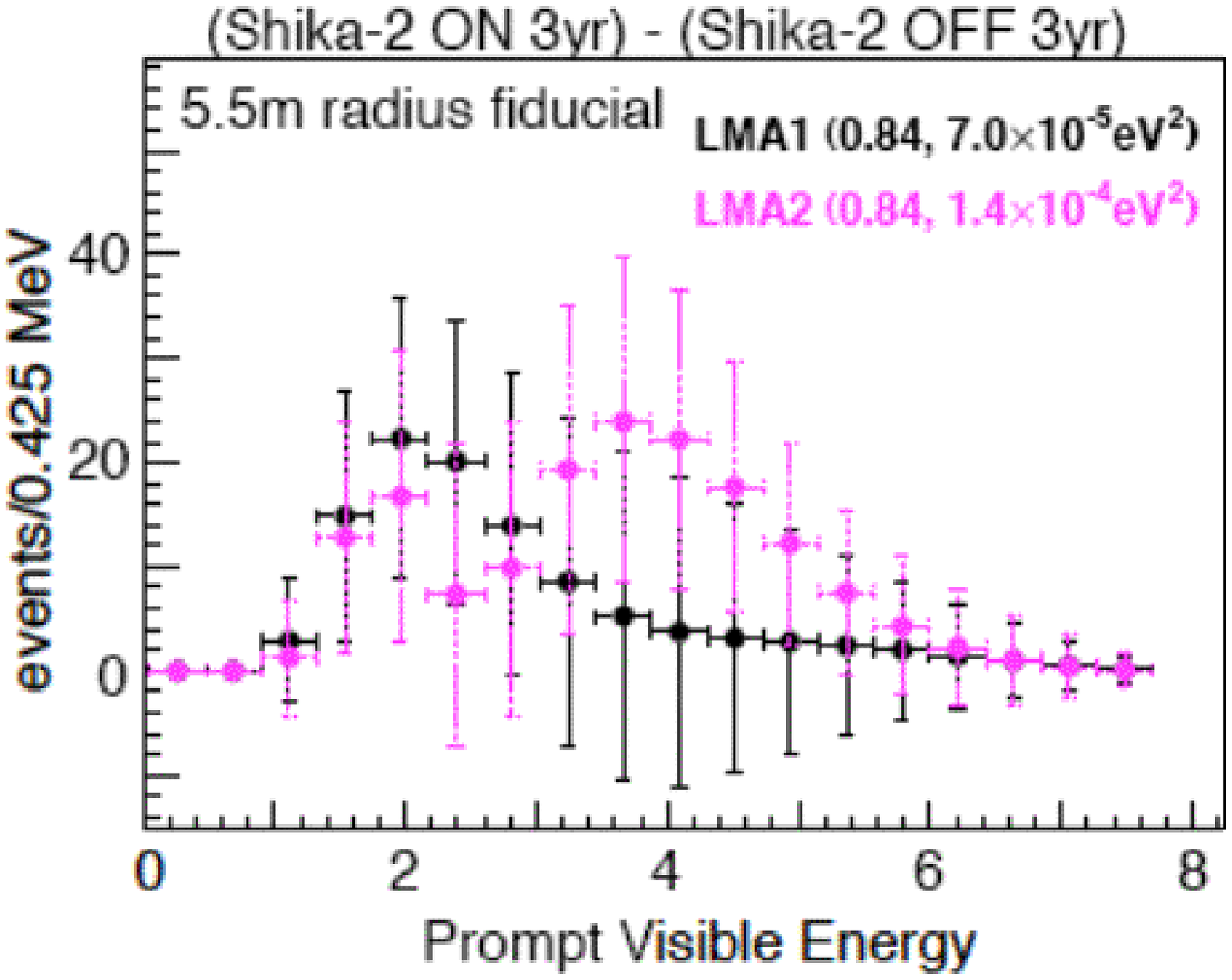,height=6.4cm}
\vspace{-2mm}
\caption{Expected $\nuebar$ spectrum from the new reactor ``Shika2''after
3 years exposure, considering the neutrino oscillation solutions LMA-1 and
LMA-2.}
\label{fig-shika-spectrum}
\end{center}
\vspace{-5mm}
\end{figure}

KamLAND is taking data continuously and an update on the reactor
$\nuebar$ analysis will be coming soon.
The expected $\nuebar$ flux is changing because 
some reactor power plants have shutdown for maintenance from
September 2002. Figure \ref{fig-modulation} shows the time variation of
the expected $\nuebar$ flux from reactors and the observed number of $\nuebar$
events at KamLAND. It is expected that a flux modulation analysis will
check the consistency of the reactor $\nuebar$ deficit.
From 2006 another strong reactor ``Shika2'' located
88km from KamLAND will begin operation.
Figure \ref{fig-shika-spectrum} shows the contribution to the
$\nuebar$ spectrum from ``Shika2''
considering the neutrino oscillation solutions LMA-1 and
LMA-2. Since the distance 88km is the most sensitive region to distinguish
LMA-1 and LMA-2, it is expected that spectrum shape analysis will 
exclude one of these two solutions.
\vspace{-3mm}

\vspace{-2mm}
\section{Result of Solar $\nuebar$ Search}

Although neutrino oscillation is the most favored solution to explain
the solar neutrino deficit other possible solutions are not completely
excluded. The search for solar $\nuebar$ \cite{kl-2nd}$^,$\,\cite{ogawa-phd}
is an effective test to examine
other exotic solutions. In this section an analysis 
of 185.5 live days KamLAND data is discussed.
\vspace{-2mm}
\subsection{Possible Mechanisms to Allow the Solar $\nuebar$}

It is generally believed that neutrinos from the Sun are generated by fusion
of light nuclei (mainly protons) in the core.
Anti neutrinos are not generated among the many processes of the fusion
in the Sun, so we should assume some other mechanism to produce solar 
$\nuebar$.
In this letter two models are discussed.
The first is a hybrid model \cite{rsfp-msw} of
resident spin flavor precession (RSFP) and the
Mikheyev-Smirnov-Wolfenstein (MSW) effect.
In this model $\nu_e$ with non-zero transition magnetic moment can evolve
into $\bar{\nu_\mu}$ or $\bar{\nu_\tau}$ while propagating through intense
magnetic fields in the solar core and they can in turn evolve into
$\nuebar$ via the MSW effect.
The other mechanism comes from a model of neutrino decay  \cite{nudecay},
where a heavy neutrino mass eigenstate may decay into a lighter
anti-neutrino mass eigenstate.

\subsection{Event Selection for Solar $\nuebar$}

\begin{figure}[h]
\begin{center}
\epsfig{figure=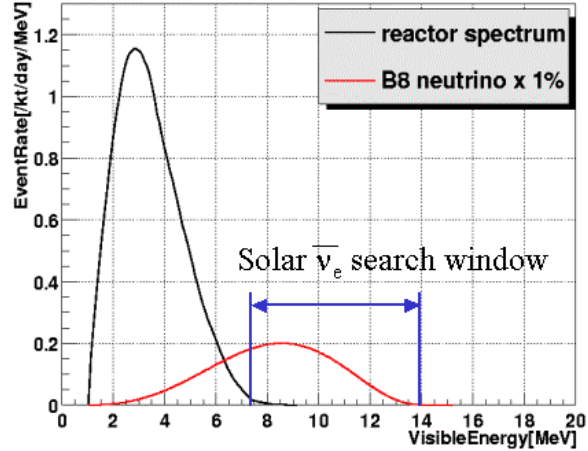,height=6.2cm}
\vspace{-2mm}
\caption{Expected visible energy spectrum of reactor $\nuebar$ together with
the $\nuebar$ spectrum from $^8$B neutrino assuming that 1\%
of flux is converted from $\nu_e$ to $\nuebar$.
The analysis energy range of solar $\nuebar$ was 7.5MeV$< E_{prompt} <$14.0MeV.
}
\label{fig-solar-nuebar-window}
\end{center}
\end{figure}

\begin{figure}[h]
\begin{center}
\epsfig{figure=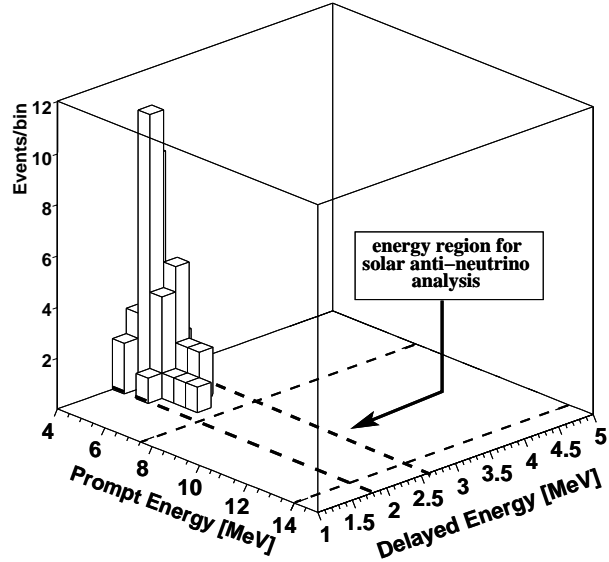,height=7.5cm}
\vspace{-2mm}
\caption{Prompt and delayed energy distribution from the solar $\nuebar$
candidates before the final energy cut (7) (see text).
No events remained in the energy range
7.5MeV$< E_{prompt} <$14.0MeV and 1.8MeV$< E_{delay} <$2.6MeV.
}
\label{fig-solar-nuebar-legoplot}
\end{center}
\end{figure}

The dominant component of solar $\nu_e$ flux above the current KamLAND
analysis threshold (2.6MeV) is the $^8$B neutrino flux, which extends
up to 14 MeV.
The reactor $\nuebar$'s, who's energies extend up to 7 MeV,
become a source of background events for the solar $\nuebar$ analysis.
Figure \ref{fig-solar-nuebar-window} shows
expected visible energy spectrum of the reactor $\nuebar$ together with
the $\nuebar$ spectrum from $^8$B neutrino assuming that 1\%
of flux is converted from $\nu_e$ to $\nuebar$.
The lower analysis threshold (7.5MeV) corresponds to the end point of reactor
$\nuebar$ spectrum and upper one (14.0MeV) corresponds to that of the
$^8$B $\nu_e$ spectrum. 
The criteria to select solar $\nuebar$ events is
(1) total charge cut, less than 10,000 p.e.
($\sim$30MeV) with no OD veto signal
and muon spallation cut.
(2) fiducial volume cut ($R < 5.5$m),
(3) time correlation cut (0.5$\mu$s $< \Delta T <$ 660$\mu$s),
(4) vertex correlation cut ($\Delta R < $1.6m),
(5) delayed energy window cut (1.8MeV $< E_{delay} <$ 2.6MeV),
(6) LS thermometer cut, and finally
(7) prompt energy cut for solar $\nuebar$ (7.5MeV $< E_{prompt} <$ 14.0MeV).
Data from 185.5 live days were included in this analysis.
Figure \ref{fig-solar-nuebar-legoplot} shows the prompt and delayed
energy distribution of the candidate events before cut (7).
After cut (7) no events remained. The total expected background was estimated
to be 1.1$\pm$0.4 events,
composing the backgrounds from reactor $\nuebar$ (0.2$\pm0.2$),
atmospheric $\nu$ (0.001), fast neutrons (0.3$\pm$0.2),
accidental coincidences (0.02) and $^8$He/$^9$Li (0.6$\pm$0.2).
Systematic uncertainties are summarized in 
Table \ref{table-solar-nuebar-systematics}.
\vspace{-2mm}
\vspace{-5mm}
\begin{table}[h]
\begin{center}
\caption{Systematic uncertainties of solar $\nuebar$ (\%).}
\vspace{2mm}
{\small
\begin{tabular}{|lc|lc|} \hline
Detection Efficiency & 1.6 & Cross section & 0.2\\
Number of target protons & 4.3 & Energy threshold& 4.3\\
Live time & 0.07 &  & \\
\hline
Total systematic error & & & 6.3\\
\hline
\end{tabular}
}
\label{table-solar-nuebar-systematics}
\end{center}
\end{table}

\vspace{-4mm}
\subsection{Upper Limit of Solar $\nuebar$ Flux}

Since no candidates are found in the corresponding energy region,
the upper limit of solar $\nuebar$ flux is calculated using
the Feldman-Cousins method \cite{feldman} with
$\nuebar$ cross section ($\sigma=6.88\times10^{-42}$cm$^2$),
detection efficiency ($\varepsilon=0.841$), live time ($T=1.60\times10^7$s)
and number of target protons ($\rho_p \times f_v=4.61\times10^{31}$).
At the 90\% C.L.
the upper limit of solar $\nuebar$ flux 
is $\phi_{\nuebar} < 3.7\times10^2$ cm$^{-2}$s$^{-1}$.
Since 29.5\% of total $^8$B neutrino flux of
$5.05 ^{+1.01}_{-0.81}\times10^6$cm$^{-2}$s$^{-1}$
is contained within the energy window of this analysis,
the $\nu_e$ to $\nuebar$ conversion probability without neutrino oscillation
is $2.8\times10^{-4}$ at 90\% C.L. 
If we assume this conversion occurs by RSFP+MSW \cite{rsfp-msw}
and also assume the
recent best fit of oscillation parameters \cite{best-fit-noon04}
($\sin^2\theta=0.28, \Delta m^2=7.2\times10^{-5}$eV$^2$),
and a solar magnetic field model \cite{solar-magnetic},
the upper limit of the neutrino transition magnetic moment $\mu_\nu$ and
magnetic field $B_{max}$ is estimated as
$\mu_\nu\cdot B_{max} < 1.4\times10^{-5} \mu_B G$ (90\%C.L.).
If we assume neutrino decay,\cite{nudecay}
we can constrain the lifetime limit
to $\tau_2/m_2 > 6.7\times10^{-2}$ s/eV.

\vspace{-2mm}
\section{Summary}
\vspace{-2mm}
KamLAND has observed an evidence for the reactor $\nuebar$ disappearance
at 99.95\% C.L. Assuming CPT invariance only the LMA solution is compatible
with the deficit.
We got an upper limit of the solar
$\nu_e$ to $\nuebar$ conversion probability $2.8\times10^{-4}$
at 90\% C.L.
in the energy range 8.3MeV $< E_{\nuebar}$ 14.8MeV.
The KamLAND experiment is supported by the COE program of the Japanese
Ministry of Education, Culture, Sports, Science, and Technology
and the United States Department of Energy.


\end{document}